\documentstyle[a4,12pt]{article}
\def\be{\begin{equation}}
\def\Bpi{$B \rightarrow \pi$}
\def\ra{\rightarrow}
\def\Dpi{$D \rightarrow \pi $}
\def\vp{\varphi}
\def\al{\alpha}
\def\ee{\end{equation}}
\def\bq{\begin{eqnarray}}
\def\eq{\end{eqnarray}}
\begin{document}
\thispagestyle{empty}
\setcounter{page}{0}
\setcounter{page}{0}
\begin{flushright}
MPI-PhT/94-79\\
LMU 19/94\\
November 1994
\end{flushright}
\vspace*{\fill}
\begin{center}
{\Large\bf Heavy-to-Light Form Factors from QCD Sum Rules
on the Light-Cone}$^*$\\
\vspace{2em}
\large
Alexander Khodjamirian$^{a,b,1}$ and Reinhold R\"uckl$^{a,c,2}$\\
\vspace{2em}
{$^a$ \small Sektion Physik der Universit\"at
M\"unchen, D-80333 M\"unchen, Germany }\\
{$^b$ \small Yerevan Physics Institute, 375036 Yerevan, Armenia } \\
{$^c$ \small Max-Planck-Institut f\"ur Physik, Werner-Heisenberg-Institut,
D-80805 M\"unchen, Germany}\\

\end{center}
\vspace*{\fill}

\begin{abstract}
We discuss the calculation of the
\Bpi~ and \Dpi~ form factors based on
an expansion in terms of pion  wave functions  on the
light-cone with increasing twist and QCD sum rule methods.
The results are compared with
predictions of conventional QCD sum rules and other approaches.
\end{abstract}
\vspace*{\fill}

\begin{flushleft}
\noindent$^1$ Alexander von Humboldt Fellow\\
\noindent$^2$ supported in part by the German Federal Ministry for Research and
Technology (BMFT) under contract No. 05 6MU93P\\
\noindent$^*$
{\it presented by A. Khodjamirian at the Workshop
"Theory of Hadrons and Light-Front QCD", Zakopane,
Poland, August 1994 }
\baselineskip=16pt
\end{flushleft}

\newpage
\section{Introduction}
The method of
QCD sum rules $^1$ has proved to be particularly useful in heavy quark
physics, where a small distance scale is provided
by the inverse heavy quark mass.
In order to calculate form factors of heavy hadrons
one can consider suitable three-point
vacuum correlation functions and apply the
operator product expansion in terms of vacuum condensates
which take into account nonperturbative quark-gluon dynamics .
Some recent calculations along  these lines can be found
in refs. $^{2,3}$.
In this report we present an alternative, more economical method
based on an expansion of vacuum-to-pion matrix elements
near the light-cone. This method is used in ref. $^{4}$ to
calculate the \Bpi~ and \Dpi~ form factors.
After outlining the calculational
procedure we show numerical results and
discuss their sensitivity to various input parameters. We also
compare our predictions with other estimates.

\section{Derivation of the sum rules}

For definiteness, we focus on the $D \rightarrow \pi$
form factor $f^{+}_{D}$ entering the matrix element
\be
<\pi|\bar{d} \gamma_\mu c |D>= 2f_D^+(p^2)q_\mu +
[f_D^+(p^2)+f_D^-(p^2)]p_\mu
\label{form}
\ee
with $p+q$, $q$ and $p$ being the $D$ and $\pi$ momenta and the
momentum transfer, respectively.
The corresponding $B \rightarrow \pi$ form factor $f^+_B $ can be
treated in parallel by obvious formal replacements.
These form factors are measurable in $B,D\rightarrow \pi l \nu_l $
semileptonic decays.

Let us consider the matrix element
\be
F_\mu (p,q)=
i \int d^4xe^{ipx}\langle \pi(q)\mid T\{\bar{d}(x)\gamma_\mu c(x),
\bar{c}(0)i\gamma_5 u(0)\}\mid 0\rangle
\label{1a}
\ee
$$
= F((p+q)^2,p^2) q_\mu + \tilde{F}((p+q)^2,p^2) p_\mu
$$
between the vacuum and an on-shell pion state.
This object is represented diagrammatically in
Fig. 1. The pion momentum squared, $q^2 = m_\pi^2$, vanishes in the
chiral limit adopted throughout this discussion.
Then, contracting the $c$-quark fields in (\ref{1a}) and keeping
only the lowest order term, i.e. the free $c$-quark propagator,
yields
\be
F ((p+q)^2,p^2)=
i \int d^4x \int \frac{d^4P}{(2\pi)^4}e^{i(p-P)x}
\sum_a \frac{\phi_a(x^2,q \cdot x)}{P^2-m_c^2}~,
\label{F}
\ee
where
\be
\phi_a(x^2, q\cdot x) =
<\pi(q)\mid \bar{d}(x)\Gamma_a u(0)\mid 0>~,
\label{phi}
\ee
$\Gamma_a$ denoting certain combinations of Dirac matrices. This
approximation corresponds to Fig. 1a.

If $(p+q)^2$ is taken sufficiently large and negative,
and the time-like momentum transfer squared  $p^2$ is far from the
kinematical limit, $p^2 = m_D^2 $,
the $c$-quark propagating between the points $x$ and $0$
is far off-shell.
 In that case, it is
justified to keep only the first few terms in the expansion
of the matrix elements (\ref{phi}) around $x^2=0$, that is near
the light-cone:
\be
\phi_a(x^2, q \cdot x) = \sum_n
\int_0^1 du\varphi_a^n(u)exp(iuq\cdot x)~ (x^2)^n
\label{phi1}~.
\ee
The form of the expansion is dictated by translational invariance.
Logarithms in $x^2$ which may also appear in (\ref{phi1}) are
disregarded for simplicity. These terms can
be consistently treated by means of QCD perturbation theory.
They give rise to normalization scale dependence.
Inserting (\ref{phi1}) into (\ref{F}) and
integrating over $x$ and $P$, one obtains, schematically,
\be
F((p+q)^2, p^2) =
\sum_a \sum_{n } \int^1_0 du \frac{\varphi^n_a(u)}
{[m_c^2-(p+qu)^2]^{2n}}~.
\label{F1}
\ee
It is thus possible to
calculate the invariant function $F$ with reasonable accuracy in the
kinematical region of highly virtual $c$-quarks provided one
knows the distribution functions $\varphi_a^n(u)$
at least for low values of $n$. The latter
are nothing but the light-cone wave functions of the
pion introduced in the context of hard exclusive
processes $^{5-7}$.

The leading twist 2 wave function is defined by
\be
< \pi(q) \mid \bar{d}(x) \gamma_\mu \gamma_5 P exp
\{ i \int ^1_0d\alpha~ x_\mu A^\mu (\alpha x) \} u(0) \mid 0 >=
-if_\pi q_\mu \int ^1_0 du e^{iuq \cdot x} \varphi_\pi(u)~,
\label{tw2}
\ee
where the exponential factor involving the gluon field is necessary
for gauge invariance.
The asymptotic form of $\varphi_\pi$  is well known:
$ \varphi_\pi(u)= 6u(1-u) $.
In our calculation of $f^+_D$ and  $f^+_B$ we have included
quark-antiquark wave functions up to twist four. In addition, we
have also calculated the first-order correction to the free
$b$-quark propagation shown in Fig. 1b which involves
quark-antiquark-gluon wave functions of twist 3 and four.
On the other hand, the  perturbative $O(\alpha_s)$ corrections
corresponding to inserting gluon exchanges between quark lines
in Fig. 1a have not been evaluated directly but have only been taken
into account in a rough indirect way, as explained below.

In order to extract the desired form factor $f^+_D$ from the result
on the invariant function $F((p+q)^2,p^2)$ sketched in (\ref{F1})
we employ a QCD sum rule with respect to the $D$-meson channel.
Writing  a dispersion relation in $(p+q)^2$, we approximate the
hadronic spectral function in the $D$-channel by the pole
contribution of the $D$ meson and a continuum contribution.
In accordance with quark-hadron duality, the latter is
identified with the spectral function derived from the QCD
representation (\ref{F1}) above the threshold $(p+q)^2=s_c$.
Formally, subtraction of the continuum then amounts to simply
changing the lower integration boundary in (7) from 0 to
$\Delta = (m_c^2-p^2)/(s_c-p^2) $. After Borel transformation
one arrives at a sum rule for the product $f_Df_D^+$, where
$f_D$ is the $D$ meson decay constant:
$$
f_Df^+_D( p^2)= \frac{f_\pi m_c^2}{2m_D^2}
\Bigg \{\int_\Delta^1\frac{du}{u}
\exp\left[\frac{m_D^2}{M^2}-\frac{m_c^2-p^2(1-u)}{uM^2}\right]
\Phi_2(u,M^2,p^2)
$$
$$
-\int_0^1\!\!u du\!\int_0^1 d\al_1\int_0^{1-\al_1} d\al_2
\frac{
\Theta( \alpha_1+u\alpha_2-\Delta)}{(\alpha_1+u\alpha_2)^2}
$$
\be
\times \exp\!\left[\frac{m_D^2}{M^2}-\frac{m_c^2-p^2
(1-\alpha_1-u\alpha_2)}{(\alpha_1+
u\alpha_2)M^2}\right]\!\Phi_3(u,\alpha_1,\alpha_2,
M^2,p^2) \Bigg\}\nonumber~,
\label{formSR}
\ee
where
\be
\Phi_2 = \vp_\pi(u) +
\frac{\mu_\pi}{m_c}\Bigg[u \vp _{p}(u)
+ \frac16 \vp_{\sigma }(u)
\left(2 + \frac{m_c^2+p^2}{uM^2}\right)\Bigg]+...~,
\label{phi2}
\ee
\be
\Phi_3= \frac{2f_{3\pi}}{f_{\pi}m_c}
\varphi_{3\pi}(\al_1,1-\al_1-\al_2,\al_2)
\left[1-\frac{ m^2_c -p^2 }{(\alpha_1+u\alpha_2)M^2}\right] +...
\label{phi3}
\ee
Here, $\varphi_p$, $\varphi_\sigma$, and $\varphi_{3\pi}$ represent twist 3
pion wave functions, while the ellipses denote contributions of higher
twist. The contributions of twist 4 are given explicitly in refs. $^{4,8}$.
The analogous sum rule  for the  $B \ra \pi$ form factor follows from
the above by formally changing $c \ra b $ and $D \ra \bar B $.

\section{Numerical evaluation}

The numerical values to be substituted
for $m_c$, $f_D$ and the threshold $s_c$ are interrelated by the
QCD sum rule for the two-point correlation
function  $\langle$ 0 $\mid T \{ j_5(x),j^+_5(0)\}\mid$ 0 $\rangle$,
$j_5= \bar{c}i\gamma_5 u$.
This sum rule should be used without
$O(\alpha_s)$ corrections in order to be consistent with the
neglect of these corrections in the sum rule for $f_Df_D^+$
given above. A similar interrelation exists for $m_b$, $f_B$ and $s_b$ from
the analogous correlation function of $b-$ flavoured currents.
For the wave functions we use the parametrization  suggested in ref. $^9$.
Other details on the choice of parameters are given in
refs. $^{4,8}$.

The form factor $f^+_D$ derived from (\ref{formSR}) is plotted
in Fig. 2 as a function of Borel mass squared $M^2$.
Numerically, the twist 3 contributions turn out to be
more important than the nonasymptotic corrections to the leading
twist 2 wave function. In the range 3 $~< M^2 <~ $ 5 GeV$^2$
the corrections due to
twist 4 in (\ref{formSR}) remain subdominant and,
simultaneously, the contribution from excited and continuum states
 does not exceed $30\%$. Restricting
oneself to this interval, one obtains the value
$f^+_D(0)= 0.66\pm 0.03$ for the $D \rightarrow \pi$ form factor
at zero momentum transfer.
The analogous fiducial interval for the $B \rightarrow \pi $ form factor
is 8 $< ~ M^2 ~ <$ 12 GeV$^2$ yielding  $f^+_B(0)= 0.29\pm 0.01~ $.

The maximum momentum transfer $p^2$ to which
these sum rules are applicable is estimated to be about
1 GeV$^2$  for $f^+_D$ and about 15 GeV$^2 $ for $f^+_B$.
The $p^2-$dependence  of both form factors is plotted
in Fig. 3.
It is important to investigate the theoretical
uncertainties in these results. Two main sources are the nonasymptotic
corrections to the leading twist wave function $\varphi_\pi$ and to the
twist 3 three-particle wave function $\varphi_{3\pi}$. In order to
estimate the sensitivity of our results to these corrections we drop them
and recalculate the form factors. As can be seen from Fig. 3,
the result changes by less than 10 \%.

\section{Conclusion}

Summarizing our investigations, in Fig. 4 we compare our predictions on
$f^+_D(p^2)$ and $f^+_{B}(p^2)$ with the results of other
calculations. Within the uncertainties there is satisfactory agreement.
In particular, we would like to
emphasize the coincidence with the result
$f^+_B(0) = 0.24 \pm 0.025 $ derived from conventional
QCD sum rule $^2$ in which the large-distance effects are parametrized in
terms of vacuum condensates rather than by pion wave functions
on the light-cone. On the other hand, the value
$f^+_D(0) = 0.5 \pm 0.1 $  obtained in ref. $^3$ is smaller than
ours.
Also the $p^2$-dependence of the form factors is rather similar in the
different approaches. Note, however, that in the quark model $^{10}$
the momentum dependence of the form factors is not predicted but simply
assumed to  be given by a single pole:
\be
f^+(p^2) = \frac{f^+(0)}{1-p^2/m_*^2}
\label{21a}
\ee
with $m_*=$ 2.01 GeV in the case of $f^+_D$ and $m_*= $5.3 GeV for
$ f^+_B$ as expected in the spirit of vector dominance. The authors of
refs. $^{2,3}$ have fitted their calculated shape for $f^+$
to the form (\ref{21a}) and obtained
$m_* = 1.95 \pm 0.10$ GeV for $f^+_D$ and
$m_* = 5.2 \pm 0.05$ GeV for $f^+_B$.
In comparison to that we find a somewhat steeper $p^2$-dependence.

In conclusion, we emphasize that light-cone sum rules such as the
ones exemplified in this report represent a well defined
alternative to the conventional QCD sum rule method. In this
variant, the nonperturbative aspects are described by a set of
wave functions on the light-cone with varying twist and quark-gluon
multiplicity.
These universal functions can be studied in a variety of processes
involving the $\pi$ and $K$ meson, or other light mesons.
The main problem to be solved if one wants to fully exploit the
light-cone approach is the reliable determination of the
nonasymptotic effects in the wave functions. In this respect,
measurements of hadronic form factors, couplings etc. can
provide important information.
A second, mainly technical problem, concerns the higher order
perturbative corrections which are still unknown.

The most important advantage of the light-cone sum rules is the
possibility to take light hadrons on mass-shell from the
very beginning.
One thus avoids the notorious model-dependence of extrapolations from
Euclidean to physical momenta in light channels.
Furthermore, in many cases the light-cone approach
is technically much easier than a conventional
QCD sum rule calculation. Finally, the light-cone method is rather
versatile. It can also be profitably employed to calculate
heavy-to-light form factors such as $B \rightarrow \rho$ and
$B \rightarrow K^*$, amplitudes of rare decays $^{11}$
and hadronic couplings such as $D^*D\pi$ and $B^*B \pi$ $^8$ .

\vskip.4in
{\Large \bf Acknowledgements}
\vskip.25in
A.K. is grateful to
the Institute of Theoretical Physics of the Warsaw
University and to the A. von Humboldt Foundation for
supporting his participation in the workshop. Useful discussions
with D. Wyler are acknowledged.
\vskip.4in

{\Large\bf References}
\vskip.25in

1. M.A. Shifman, A.I. Vainshtein, V.I. Zakharov, Nucl.
Phys. {\bf B147} (1979) 385, 448.

2. P. Ball, V.M. Braun, H.G. Dosch,
Phys. Lett. {\bf B273} (1991) 316 .

3. P. Ball, Phys. Rev. {\bf D48} (1993) 3190.

4. V. Belyaev, A. Khodjamirian, R. R\"uckl,
Z. Phys. {\bf C60} (1993) 349.

5. G.P. Lepage and S.J. Brodsky,
  Phys. Lett.  {\bf {B87}} (1979) 359;
  Phys. Rev.  {\bf {D22}}

(1980) 2157.

6. A.V. Efremov and A.V. Radyushkin,
  Phys. Lett.  {\bf {B94}} (1980) 245.

7. V.L. Chernyak, A.R. Zhitnisky,
Phys. Reports {\bf 112} (1984) 173.

8. V.M. Belyaev, V.M. Braun, A. Khodjamirian,
R. R\"uckl, preprint MPI- PhT/94-62.

9. V.M. Braun, I. Filyanov,
Z. Phys. {\bf C44} (1989) 157; {\it ibid.}
{\bf C48} (1990) 239.

10. M. Bauer, B. Stech, M. Wirbel, Z. Phys. {\bf C34} (1987) 103.

11. A. Ali, V.M. Braun, H. Simma, Z. Phys. {\bf C63} (1994) 437.

\end{document}